\documentclass[12pt,preprint]{aastex}

\def\hal{H$\alpha$}
\def\hbeta{H$\beta$}
\def\be{\begin{equation}}
\def\ee{\end{equation}}

\def\m{~$\mu$m}

\def\NII  {[\ion{N}{2}]}
\def\NII  {[\ion{N}{2}]}

\def\OII  {[\ion{O}{2}]}
\def\OIII {[\ion{O}{3}]}

\def\Nsourcesa{214}
\def\Nsourcesb{424}
\def\Nsourcesc{438}
\def\Nsourcesd{91}

\def\rhoa{0.010}
\def\rhob{0.013}
\def\rhoc{0.020}
\def\rhod{0.022}

\begin {document}

\title{The Wyoming Survey for H$\alpha$.  III. H$\alpha$ Luminosity Functions at $z\approx$~0.16, 0.24, 0.32, and 0.40}
\shorttitle{The Wyoming Survey for H$\alpha$}

\author {Daniel~A.~Dale\altaffilmark{1}, Rebecca~J.~Barlow\altaffilmark{2}, Seth~A.~Cohen\altaffilmark{3}, David~O.~Cook\altaffilmark{1}, L.~Clifton~Johnson\altaffilmark{4}, ShiAnne~M.~Kattner\altaffilmark{5}, Carolynn~A.~Moore\altaffilmark{1}, Micah~D.~Schuster\altaffilmark{6}, Shawn~M.~Staudaher\altaffilmark{1}}
\altaffiltext{1} {\scriptsize Department of Physics and Astronomy, University of Wyoming, Laramie, WY 82071}
\altaffiltext{2} {\scriptsize Department of Physics, University of New Hampshire, Durham, NH 03824}
\altaffiltext{3} {\scriptsize Department of Physics and Astronomy, Dartmouth University, Hanover, NH 03747}
\altaffiltext{4} {\scriptsize Department of Astronomy, University of Washington, Seattle, WA 98195}
\altaffiltext{5} {\scriptsize Department of Astronomy, San Diego State University, San Diego, CA 92182}
\altaffiltext{6} {\scriptsize Department of Physics, San Diego State University, San Diego, CA 92182}

\begin {abstract}
The {\it Wyoming Survey for H$\alpha$}, or WySH, is a large-area, ground-based imaging survey for \hal-emitting galaxies at redshifts of $z\approx0.16$, 0.24, 0.32, and 0.40.  The survey spans up to four square degrees in a set of fields of low Galactic cirrus emission, using twin narrowband filters at each epoch for improved stellar continuum subtraction.  H$\alpha$ luminosity functions are presented for each $\Delta z \approx 0.02$ epoch based on a total of nearly 1200 galaxies.  These data clearly show an evolution with lookback time in the volume-averaged cosmic star formation rate.  Integrals of Schechter fits to the incompleteness- and extinction-corrected \hal\ luminosity functions indicate star formation rates per co-moving volume of \rhoa, \rhob, \rhoc, \rhod ~$h_{70}~M_\odot$~yr$^{-1}$~Mpc$^{-3}$ at $z \sim 0.16$, 0.24, 0.32, and 0.40, respectively.  Statistical and systematic measurement uncertainties combined are on the order of 25\% while the effects of cosmic variance are at the 20\% level.  The bulk of this evolution is driven by changes in the characteristic luminosity $L_*$ of the \hal\ luminosity functions, with $L_*$ for the earlier two epochs being a factor of two larger than $L_*$ at the latter two epochs; it is more difficult with this data set to decipher systematic evolutionary differences in the luminosity function amplitude and faint-end slope.  Coupling these results with a comprehensive compilation of results from the literature on emission line surveys, the evolution in the cosmic star formation rate density over $0 \lesssim z \lesssim 1.5$ is measured to be $\dot{\rho}_{\rm SFR}(z) = \dot{\rho}_{\rm SFR}(0) (1+z)^{3.4\pm0.4}$.
\end {abstract}

 
\section {Introduction}

The rate at which stars form on a cosmic scale is a fundamental property of the universe.  Various efforts have shown the cosmic star formation rate per unit co-moving volume to significantly rise from the current epoch back to a redshift of 1, evolving at the rate of $\dot{\rho}_{\rm SFR}(z) \propto (1+z)^\gamma$, where $\gamma$ is estimated to lie between 1.5 and 4 (e.g., Lilly et al.\ 1996; Cowie et al.\ 1999; Wilson et al.\ 2002; Gallego et al.\ 2002; Tresse et al.\ 2002; Hippelein et al.\ 2003; Hopkins 2004; P\'erez-Gonz\'alez et al.\ 2005; Le Floc'h et al.\ 2005; Hanish et al.\ 2006; Babbedge et al.\ 2006; Villar et al.\ 2008; Magnelli et al.\ 2009).  Reliably constraining $\gamma$ will assist in interpreting models of galaxy evolution, since the cosmic star formation history is intimately tied to the build-up of heavy metals and stellar mass over time (Pei \& Fall 1995; Madau et al.\ 1996; Pei et al.\ 1999; Somerville et al.\ 2001; Cole et al.\ 2001; Panter et al.\ 2003).  In fact, predictions of the accumulated stellar mass based on the integrated cosmic star formation history exceed the observed stellar mass at the present epoch by a factor of two (Figure 1 of Wilkins 2008).  From an observational viewpoint, the evolutionary parameter $\gamma$ is ideally measured by uniformly sampling the star formation rate using a large sample of galaxies spanning multiple fields and multiple epochs between $z=0$ and $z\sim1-2$.  The Wyoming Survey for \hal, or WySH, surveys hundreds of galaxies at four intermediate redshifts, covering three separate fields that total some four square degrees (Dale et al.\ 2008; hereafter D08).  WySH utilizes \hal\ to probe the star formation in galaxies, an optical emission line that directly traces massive star formation, is technically simple to observe, and lies at a relatively long wavelength to help minimize the effects of extinction internal to star-forming galaxies.  The survey has been carried out where redshifted \hal\ can be detected at wavelengths over which CCDs are sensitive: $z\approx0.16$, 0.24, 0.32, and 0.40.  This optical observing program is complemented by NewH$\alpha$, a collaborative near-infrared narrowband imaging survey of \hal\ at redshifts of $z \sim 0.81$ and 2.2 using the wide-field NEWFIRM camera (Probst et al.\ 2004) on the Kitt Peak National Observatory 4~m telescope (Lee et al.\ 2010; Ly et al.\ 2010).  The NewH$\alpha$ survey includes observations through low airglow windows at 1.19 and 2.10\m, sampling the \hal\ luminosity function down to levels similar to that explored in WySH.  NewH$\alpha$ also has an extensive follow-up multi-slit spectroscopy campaign with the 6.5~m Baade telescope at Las Campanas Observatory, allowing us to verify \hal-emitting candidates, directly measure internal extinction via Balmer line ratios, probe metal abundances, and to identify AGN (Momcheva et al.\ 2010).  The combination of these two surveys will provide powerful constraints on the evolution in the \hal\ luminosity function and the cosmic star formation rate over $0 \lesssim z \lesssim 2.2$.  This paper summarizes results from the WySH observing program, including \hal\ luminosity functions at four intermediate redshifts, a measure of the evolution in the volume-averaged cosmic star formation rate over these same epochs, and an estimate of the impact of cosmic variance.  Section~\ref{sec:survey} reviews the survey parameters, \S~\ref{sec:results} presents results from all epochs, and \S~\ref{sec:summary} provides a summary.  We assume $H_0=70~h_{70}$~km~s$^{-1}$~Mpc$^{-1}$, $\Omega_{\rm m}=0.3$, and $\Omega_\lambda=0.7$.

\section{The Survey}
\label{sec:survey}

WySH is a multi-epoch narrowband imaging survey for redshifted \hal\ carried out on the Wyoming Infrared Observatory 2.3~m telescope (WIRO).  The survey spans multiple ``blank'' fields selected to minimize foreground contamination by zodiacal and Galactic dust, bright stars, and nearby galaxies.  Observing multiple fields also aids in minimizing the impact of cosmic variance, a catchall term for the statistical fluctuations inherent to observations of different regions at cosmological distances.  By design, two of these fields, ELAIS-N1 and the Lockman Hole, overlap with the target areas of deep infrared and ultraviolet surveys.  The SWIRE Spitzer Legacy project provides maps of these regions using several bandpasses between 3 and 160\m\ (Lonsdale et al.\ 2003) along with a wealth of ancillary data at other wavelengths.  The GALEX Deep Imaging Survey (Martin et al.\ 2005) includes the ELAIS-N1 and Lockman Hole fields, providing two channels of ultraviolet data from integrations that are 300 times longer than the standard integration of that mission's all-sky survey.  The third WySH field, also known as ``WySH~1'', conveniently fills the right ascension gap between the Lockman Hole and the ELAIS-N1 field to enable year-round observations from WIRO (D08 Table~1).  While the main goal of this survey is to provide a statistically robust measure of the evolving star formation history of the Universe at intermediate redshifts, the combination of the surveys described above enables interesting parallel science.  For example, combining \hal, ultraviolet, and infrared data allow us to study the evolution over time of the average attenuation by interstellar dust within star-forming galaxies (Moore et al.\ 2010). 
 
\subsection{Observations and Data Processing}
\label{sec:observations}

WIRO is equipped with a prime focus CCD camera with 0\farcs523 pixels and a 17\farcm9 field of view (Pierce \& Nations 2002).  Redshifted \hal\ emission is observed in four separate cosmological epochs, $z\approx0.16, 0.24, 0.32$, and 0.40.  This survey differs from traditional narrowband imaging surveys by using nearly identical and wavelength-adjacent narrowband filter pairs ($\sim60$\AA\ FWHM).  If a source's redshift places the \hal\ emission to appear in Filter~A, then the emission detected from Filter~B is used to subtract off the stellar continuum as detected by Filter~A.  Likewise, the reverse applies if \hal\ is found in Filter~B.  The basic unit of observation is a 300~s frame, though several are taken for each filter at each location.  After the pre-processing of individual frames (e.g., bias subtraction, flat-fielding, fringing correction; see \S~2.3 of D08), the multiple 300~s frames for a given field are aligned and stacked to create images with longer effective integrations.  The effective integration increases with increasing redshift with the aim of achieving a similar sensitivity to line luminosity at each epoch: the $z\approx$[0.16, 0.24, 0.32, 0.40] stacks have effective integrations of [1200, 3600, 6000, 9600]~s.  Since two filters are utilized at each epoch, there are in fact two stacks of these integration depths for each pointing at each epoch.  The total areal coverage for the survey is [4.19, 4.03, 4.13, 1.11] square degrees at $z\approx$[0.16, 0.24, 0.32, 0.40].

The details of how sources are identified and their photometry extracted with SExtractor (Bertin \& Arnouts 1996) is described in Dale et al.\ (2008).  Salient details include a 5\% flux calibration uncertainty, a 20\% correction for \NII\ emission within the filter bandpasses, and the application of a 3$\sigma$ cut based on the \hal+\NII\ signal-to-noise.  The 3$\sigma$ continuum sensitivity to luminosity is $\sim 10^{40}$~ergs~s$^{-1}$ (and $\sim 5\cdot10^{-17}~{\rm ergs~s}^{-1}~{\rm cm}^{-2}$ in terms of flux), with the survey at the later epochs being somewhat more sensitive and the surveys at the earlier epochs somewhat less sensitive than this value, in part due to the falling quantum efficiency of the CCD with wavelength (Figure~5 of D08).  The sensitivity to line detections is an additional factor of $\sim\sqrt{2}$ poorer since line strengths are derived from a subtraction of two narrowband luminosities.

\subsection {Removal of Contamination by Redshift Interlopers}

The inclusion of non-\hal\ line emission from galaxies outside of the target epochs will bias the \hal\ luminosity function and thus the inferred star formation rate density at each epoch; other prominent optical emission lines such as \OIII $\lambda\lambda$4959/5007, \OII $\lambda$3727, and \hbeta $\lambda$4861 could be redshifted into the filter bandpasses (e.g., Fujita et al.\ 2003).  Photometric redshifts are used to cull the contaminators.  Multiple photometric redshift catalogs are available for the fields pursued in this work: Rowan-Robinson et al.\ (2007) from SWIRE, Csabai et al.\ (2003) from SDSS, and from our own suite of deep observations from WIRO covering {\it UBVRI} (15--75~min integrations) and narrowband imaging at eight wavelengths between 7597 and 9233~nm.  WIRO-based photometric redshifts are estimated using the {\it Le Phare} code\footnote{http://www.oamp.fr/people/arnouts/LE\_PHARE.html} and a suite of spiral, elliptical, irregular, and starburst templates (the {\tt CWW\_KINNEY} templates), allowing the $V$ band extinction to vary from 0 to 3~magnitudes.  For the sake of uniformity SWIRE photometric redshifts are used where available, with SDSS and our own photometric redshifts employed to supplement the SWIRE coverage.

\section {Results}
\label{sec:results}

\subsection{\hal\ Luminosity Functions}
\begin{figure}
 \plotone{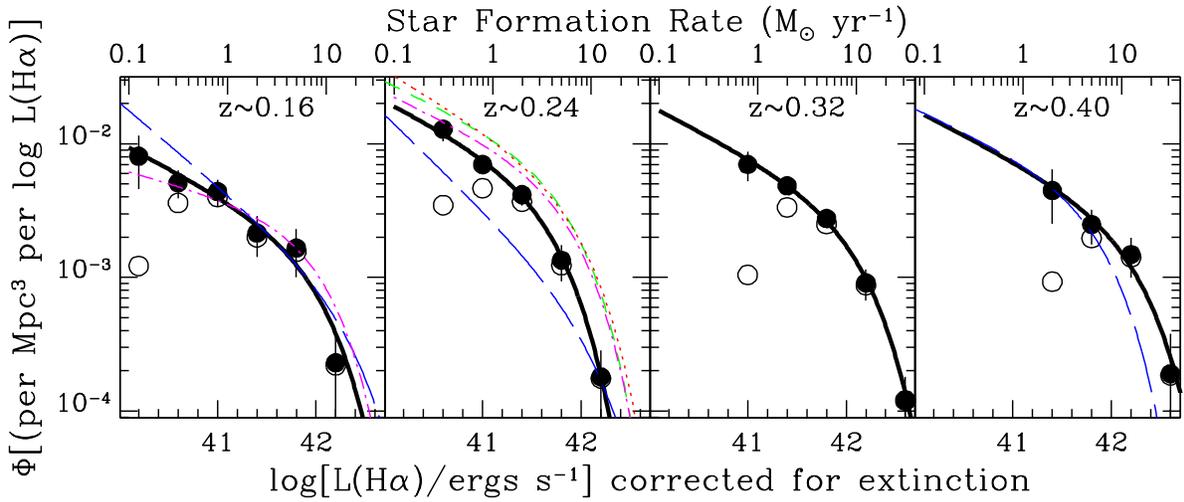}
 \caption{The luminosity functions at $z\approx 0.16$, 0.24, 0.32, and 0.40, based on \Nsourcesa +\Nsourcesb +\Nsourcesc +\Nsourcesd\ \hal-emitting galaxies.  The data without incompleteness corrections are displayed as open circles, while those corrected for incompleteness are shown as filled circles.  Error bars reflect the uncertainty in the luminosity function amplitude according to Equation~\ref{eq:lf}, summed in quadrature with the uncertainties in the incompleteness corrections.  The thick solid lines show the Schechter fits for the parameters presented in the first four rows of Table~\ref{tab:results}.  Literature fits are also provided at $z\sim0.16$ (Sullivan et al.\ 2000; Westra et al.\ 2010), $z\sim0.24$ (Tresse \& Maddox 1998; Fujita et al.\ 2003; Ly et al.\ 2007; Shioya et al. 2008), and $z\sim0.40$ (Ly et al.\ 2007).}
 \label{fig:lfs}
\end{figure}

The \hal\ luminosity function for each epoch is displayed in Figure~\ref{fig:lfs}.  The amplitude of the $i^{\rm th}$ bin of each luminosity function and its uncertainty are derived from
\be
\Phi(z,\log L_i) = {\Sigma_j V(z_j,L_j)^{-1} \over \Delta \log L}, \;\;\;\;\;\;\; \epsilon[\Phi(z,\log L_i)] = {\sqrt {\Sigma_j V(z_j,L_j)^{-2}} \over \Delta \log L}
\label{eq:lf}
\ee
where $V(z_j,L_j)$ is the co-moving volume for the $j^{\rm th}$ galaxy in the summation.  The luminosities are binned according to $\mid \log L_j - \log L_i \mid\ < \onehalf \Delta \log L$, with a luminosity bin width $\Delta \log L$ spanning 0.4~dex.  Several corrections described in \S~3 of D08 are incorporated into the luminosity functions displayed in Figure~\ref{fig:lfs}.  The luminosity functions are corrected for sample incompleteness ($\kappa(z,L)_{\rm inc}$), filter transmission characteristics, and \hal\ line extinction ($e^{\tau(L)}$) following the \hal\ luminosity-dependent prescription of Hopkins et al.\ (2001; though see also Garn et al.\ 2010).  A bin-by-bin tabulation of the incompleteness and extinction corrections are provided in Table~\ref{tab:corrections}.  Open circles in Figure~\ref{fig:lfs} indicate the data corrected for all issues described above except incompleteness; the filled circles also include corrections for incompleteness.  Error bars in Figure~\ref{fig:lfs} reflect the uncertainty in the luminosity function amplitude according to Equation~\ref{eq:lf}, summed in quadrature with the uncertainties in the incompleteness corrections.  Also included in Figure~\ref{fig:lfs} are \hal-based luminosity functions from the literature, largely consistent with our results.

\begin{figure}
 \plotone{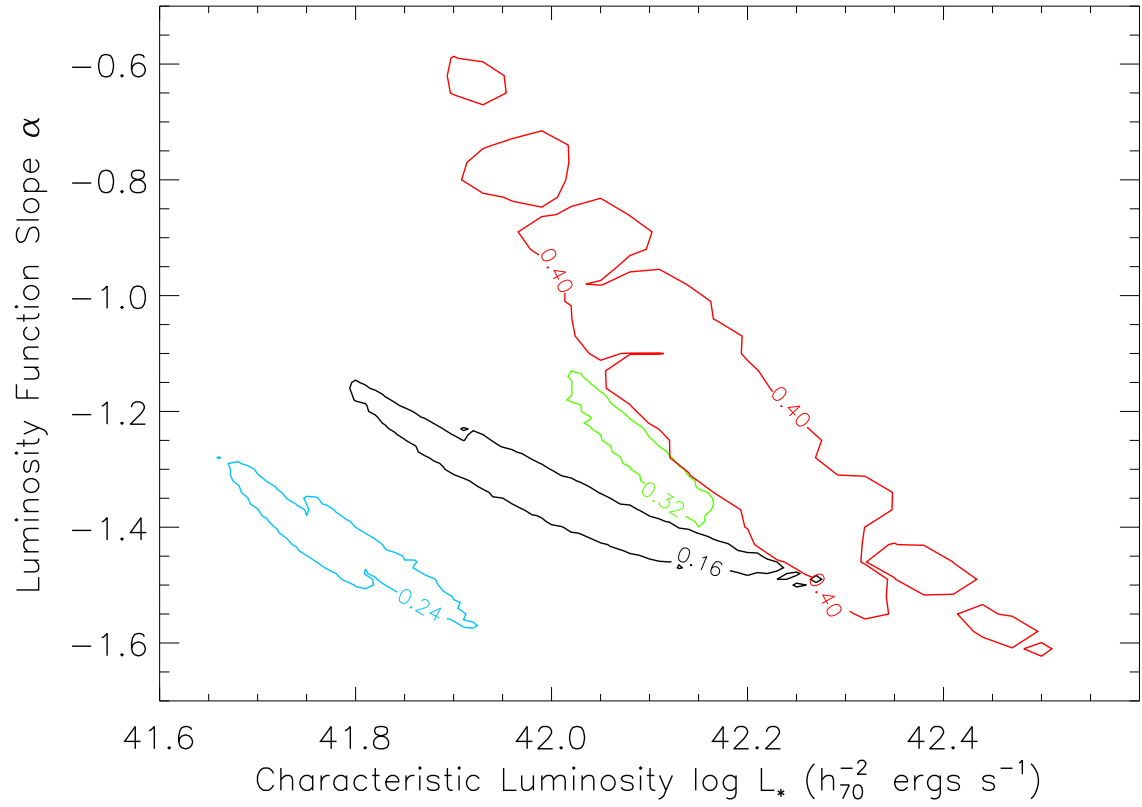}
 \caption{The 2$\sigma$ (95.4\%) confidence distributions of the luminosity function parameters $\alpha$ and $L_*$ for 10,000 Monte Carlo simulations of the data, for the case where all parameters are allowed to vary (Rows~5--8 of Table~\ref{tab:results}).  The redshifts are indicated within the contours.}
 \label{fig:monte_carlo}
\end{figure}

Going down to and including these limiting luminosities, a Schechter profile (Schechter 1976) is fit to the incompleteness-corrected luminosity functions:
\be
\Phi(z,\log L) d\log L = \phi(z,L) dL = \phi_*(z) \left( {L \over L_*(z)} \right)^{\alpha(z)} e^{-L/L*(z)} {dL \over L_*(z)},
\ee
where $\alpha(z)$ conveys the shape of the function, $L_*(z)$ sets the luminosity scale, and $\phi_*(z)$ represents the overall normalization.  The parameters for the functional fits displayed in Figure~\ref{fig:lfs} are listed in the first four rows of Table~\ref{tab:results}.  Since the faint end slopes are difficult to gauge from the available luminosity bins for the two earliest epochs, for these solutions we have fixed $\alpha(z\approx0.32)$ and $\alpha(z\approx0.40)$ to be the same as the average of that found for the later two epochs ($-1.38$).  Also included in Table~\ref{tab:results} are solutions allowing all parameters to vary.  In all cases Monte Carlo simulations are employed to estimate the average fit parameters and their uncertainties (see Figure~\ref{fig:monte_carlo}).  Each fit is repeated 10,000 times after using the measured luminosity uncertainties to randomly add a (Gaussian deviate) offset to each \hal\ luminosity.  The standard deviations in the simulations are used to represent the luminosity function parameter uncertainties.

Extremely deep surveys are required to adequately measure the faint end slopes of the luminosity function (e.g., Ly et al. 2007).  As can be seen from Figure~\ref{fig:lfs}, the lowest luminosities for which $\Phi(z,\log L)$ are reliably obtained increase with redshift: $\log L_{\rm lim}({\rm H}\alpha/{\rm ergs~s}^{-1})=$[40.2, 40.6, 41.0, 41.4] for $z\approx$ [0.16, 0.24, 0.32, 0.40].  While the fitted values for $\alpha(z)$ are somewhat steeper for $z\approx0.16$ and 0.24 than at the two earlier epochs (see Rows 5--8 in Table~\ref{tab:results}), this result is very tenuous since there are too few low luminosity galaxies detected at $z\approx0.32$ and 0.40 to adequately constrain the faint end slopes.  

In terms of the other luminosity function parameters, $L_*(z)$ and $\phi_*(z)$, Westra et al.\ (2010) show that intermediate redshift surveys like WySH that span at least 3 square degrees cover large enough volumes to effectively constrain the bright end of the luminosity function.  For our standard solutions involving a fixed slope of $\alpha=-1.38$ and luminosity-dependent internal extinction, the changes in the luminosity function with redshift are mainly driven by changes in the characteristic luminosity $L_*(z)$, with the average $L_*(z)$ being a factor of two larger for the two earlier epochs than the average measured at $z\approx0.16$ and 0.24.  It is more difficult to discern any systematic evolutionary changes in the source number density as parameterized by the luminosity function amplitude $\phi_*(z)$.

\subsection {The Evolution in the Cosmic Star Formation Density}

\begin{figure}
 \plotone{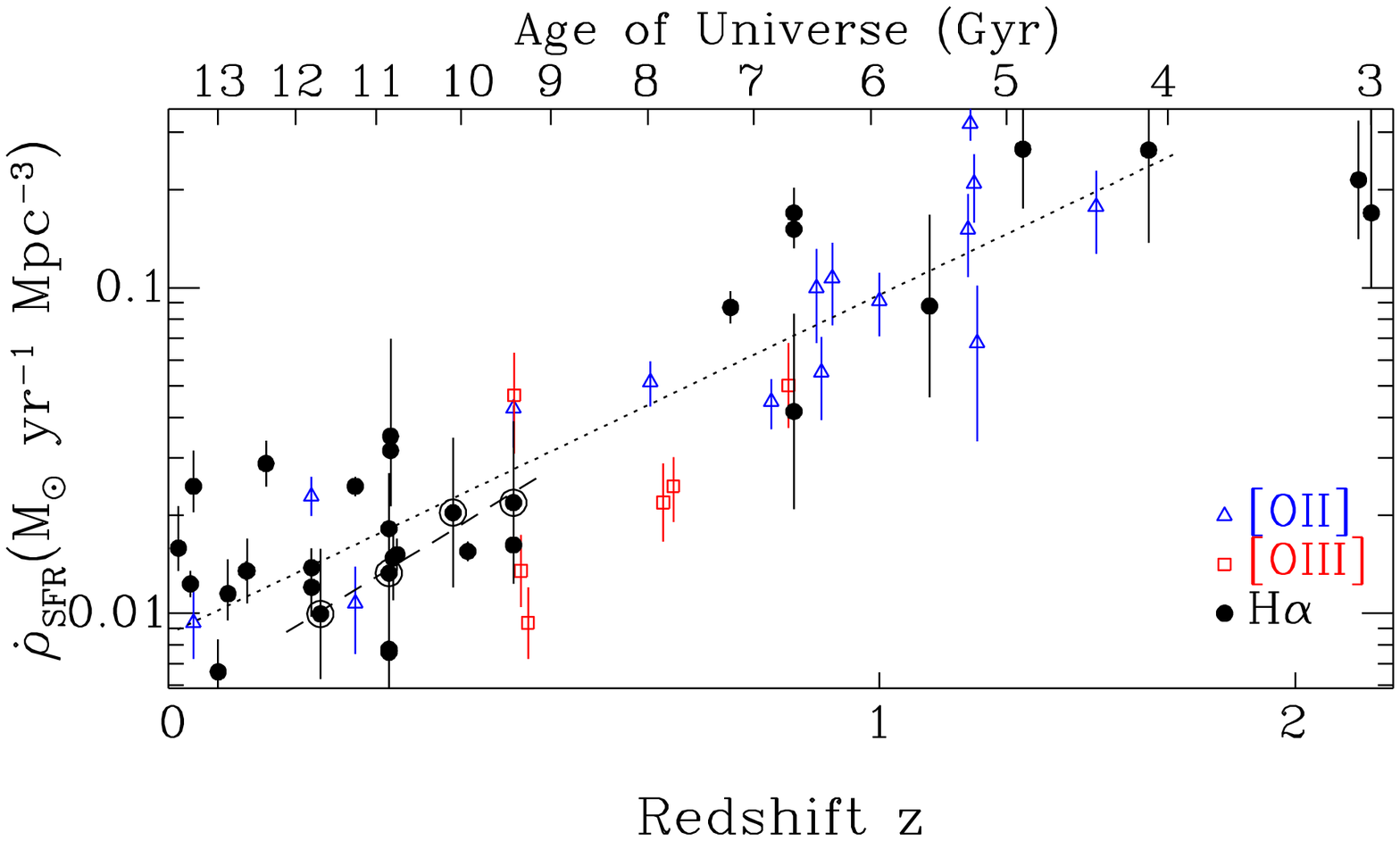}
 \caption{The cosmic star formation history, using results from line emission surveys only.  Data points from the WySH survey presented here are encircled and have a slope of $4.5\pm0.7$.  Literature data come from Gallego et al.\ (1995), Hogg et al.\ (1998), Tresse \& Maddox (1998), Yan et al.\ (1999) assuming $A_V=1$~mag, Sullivan et al.\ (2000), Gallego et al.\ (2002), Tresse et al.\ (2002), Fujita et al.\ (2003), P\'erez-Gonz\'alez et al.\ (2003), Hippelein et al.\ (2003), Nakamura et al.\ (2004), Brinchmann et al.\ (2004), Hanish et al.\ (2006), Takahashi et al.\ (2007), Ly et al.\ (2007), Geach et al.\ (2008), Westra \& Jones (2008), Morioka et al.\ (2008), Shioya et al.\ (2008), Villar at al. (2008), Dale et al.\ (2008), Shim et al.\ (2009), Sobral et al.\ (2009), Ly et al.\ (2010), Hayes et al.\ (2010), and Westra et al.\ (2010).}
 \label{fig:sfrd}
\end{figure}

The volume-averaged cosmic star formation rate can be computed by integrating under the fitted Schechter function and multiplying by the Kennicutt (1998) star formation rate calibration:
\be
\dot{\rho}_{\rm SFR}(z) (h_{70}~M_\odot~{\rm yr}^{-1}~{\rm Mpc}^{-3})= 7.9\cdot10^{-42} ~\mathcal{L}(z)(h_{70}~{\rm ergs~s}^{-1}~{\rm Mpc}^{-3})
\ee
where an analytical expression for the luminosity density is
\be
\mathcal{L}(z) = \int_0^{\infty} dL L \Phi(z,L) = \phi_{\star}(z)L_{\star}(z)\Gamma(\alpha+2).
\ee
Table~\ref{tab:results} provides the integrated cosmic star formation rate densities.  There is a clear, systematic increase in $\dot{\rho}_{\rm SFR}(z)$ with redshift, with an overall difference of a factor of $\sim2$ between $z\approx0.16$ and $z\approx0.40$.  
This result is robust to the effects of cosmic variance, fluctuations due to the characteristics of the particular volume(s) being probed along a survey's line(s)-of-sight (e.g., clusters, voids, etc.): the values for $\dot{\rho}_{\rm SFR}(z)$ extracted from individual fields (e.g., Lockman Hole and ELAIS-N1) differ by only $\sim 20$\%.  Placed in a larger context, our values for $\dot{\rho}_{\rm SFR}(z)$ fit well within the envelope determined by previous emission line surveys (see Figure~\ref{fig:sfrd} and the citations in the caption).  If the evolution over 
$0 \lesssim z \lesssim 1.5$ 
is cast in power-law form, $\dot{\rho}_{\rm SFR}(z) \propto (1+z)^\gamma$, the exponent is best fit by $\gamma=3.4\pm0.4$.  This slope is consistent with the typical values ($3 \lesssim \gamma \lesssim 4$) found in the literature for this redshift range using star formation rates based on the infrared continuum or dust-corrected optical emission lines (e.g., Gallego et al.\ 2002; Tresse et al.\ 2002; Hippelein et al.\ 2003; P\'erez-Gonz\'alez et al.\ 2005; Le Floc'h et al.\ 2005; Hanish et al.\ 2005; Babbedge et al.\ 2006; Villar et al.\ 2008; Magnelli et al.\ 2009).  Surveys based on the ultraviolet continuum, where attenuation by dust and contributions from older stellar populations can play important roles, tend to recover shallower slopes (e.g., Cowie et al.\ 1999; Wilson et al.\ 2002; Schiminovich et al.\ 2005; Prescott et al.\ 2009).  Cosmic star formation history surveys using optical emission line data at higher redshifts are understandably sparse, and Figure~\ref{fig:sfrd} provides just a hint of the leveling off of $\dot{\rho}_{\rm SFR}(z)$ beyond a redshift of $z \sim 1-1.5$ that is seen in other compilations that profit from a broader wavelength baseline (e.g., Hopkins 2004).





\section {Discussion and Summary}
\label{sec:summary}
The primary aim of the {\it Wyoming Survey for H$\alpha$}, or WySH, is to accurately quantify the \hal\ luminosity function via narrowband imaging spanning $\lesssim 4$ square degrees and multiple cosmic epochs.  Important features of the survey include the use of narrowband filter pairs at each epoch for improved stellar continuum subtraction and the spatial overlap with deep ultraviolet and infrared surveys that enable interesting follow-up studies.  Buttressed by a total of nearly 1200 \hal\ detections, we find a modest evolution in the \hal\ luminosity function over $z \sim 0.16$, 0.24, 0.32, and 0.40.  The values of the volume-averaged cosmic star formation rate, found by integrating under the luminosity functions, change by a factor of two over this moderate stretch in redshift.  Our results indicate that this evolution is largely driven by changes in the characteristic luminosity $L_*$, which also shows an evolution by a factor of two over these epochs.  That the evolution in the cosmic star formation rate density over these intermediate redshifts is mainly influenced by systematic changes in the characteristic luminosity is consistent with the findings of Le Floc'h et al.\ (2005), P\'erez-Gonz\'alez et al.\ (2005), Magnelli et al.\ (2009), and Westra et al.\ (2010), though Ly et al.\ (2007) find the evolution to be more driven by changes in the source number density.  Placing our results in the larger context of the slew of recent emission line surveys for $\dot{\rho}_{\rm SFR}$ over $0 \lesssim z \lesssim 1.5$, the evolution in the cosmic star formation rate density is estimated to be $\dot{\rho}_{\rm SFR}(z) \propto (1+z)^{3.4\pm0.4}$.  Results from a complementary near-infrared narrowband imaging survey of \hal-emitters will extend this work to redshifts of $z \sim 0.81$ and 2.2.
Finally, the large volume covered by this optical survey enables a measure of the impact of cosmic variance.  By separately analyzing the different fields in this survey, in particular the Lockman Hole and ELAIS-N1, we find a variation in $\dot{\rho}_{\rm SFR}$ at the 20\% level for any given redshift. 

\acknowledgements 
This research is funded through the NSF CAREER program (AST0348990).  
Special thanks go to Michael Pierce for developing WIROPrime and to the WIRO staff for keeping the telescope operational.  IRAF, the Image Reduction and Analysis Facility, has been developed by the National Optical Astronomy Observatories and the Space Telescope Science Institute.

\begin{deluxetable}{cccccc}
\tabletypesize{\scriptsize}
\tablenum{1}
\label{tab:corrections}
\tablecaption{H$\alpha$ Extinction and Luminosity Function Incompleteness}
\tablewidth{0pc}
\def\p{$\pm$}
\tablehead{
\colhead{log$L$(H$\alpha$)} &
\colhead{$e^{\tau(L)}$} &
\colhead{$\kappa(z,L)_{\rm inc}$}      &
\colhead{$\kappa(z,L)_{\rm inc}$}      &
\colhead{$\kappa(z,L)_{\rm inc}$}      &
\colhead{$\kappa(z,L)_{\rm inc}$}  
\\         
\colhead{(ergs s$^{-1}$)} &
\colhead{} &
\colhead{7597/7661\AA}      &
\colhead{8132/8199\AA}      &
\colhead{8615/8685\AA}      &
\colhead{9155/9233\AA}           
}
\startdata
40.2&1.18&0.15\p0.02&\nodata   &\nodata   &\nodata   \\
40.6&1.49&0.65\p0.05&0.27\p0.04&\nodata   &\nodata   \\
41.0&1.84&0.89\p0.05&0.66\p0.07&0.15\p0.05&\nodata   \\
41.4&2.24&0.91\p0.04&0.89\p0.05&0.69\p0.05&0.21\p0.05\\
41.8&2.68&0.93\p0.04&0.92\p0.04&0.91\p0.04&0.79\p0.04\\
42.2&3.17&0.96\p0.03&0.96\p0.03&0.96\p0.03&0.95\p0.03\\
42.6&3.70&\nodata   &\nodata   &0.98\p0.03&0.96\p0.03\\
\enddata
\end{deluxetable}

\begin{deluxetable}{clccc}
\tabletypesize{\scriptsize}
\tablenum{2}
\label{tab:results}
\tablecaption{Luminosity Function Results}
\tablewidth{0pc}
\def\a{\tablenotemark{a}}
\def\p{$\pm$}
\tablehead{
\colhead{Redshift}  &
\colhead{$\alpha$}&
\colhead{$\log L_*$}&
\colhead{$\log \phi_*$}&
\colhead{$\dot{\rho}_{\rm SFR}$}
\\
\colhead{}    &
\colhead{}    &
\colhead{($h_{70}^{-2}$ ergs s$^{-1}$)} &
\colhead{($h_{70}^3 {\rm Mpc}^{-3} {\rm dex}^{-1}$)}  &
\colhead{$h_{70} M_\odot {\rm yr}^{-1} {\rm Mpc}^{-3}$}
}
\startdata
0.16&$-$1.38\a    &42.0\p0.1&$-$3.08\p0.02&0.010\p0.001\\
0.24&$-$1.38\a    &41.8\p0.1&$-$2.70\p0.02&0.013\p0.001\\
0.32&$-$1.38\a    &42.2\p0.1&$-$2.90\p0.01&0.021\p0.001\\
0.40&$-$1.38\a    &42.3\p0.1&$-$2.97\p0.04&0.024\p0.002\\
\hline
0.16&$-$1.36\p0.06&42.0\p0.2&$-$3.05\p0.11&0.010\p0.001\\
0.24&$-$1.41\p0.05&41.8\p0.1&$-$2.74\p0.07&0.013\p0.001\\
0.32&$-$1.26\p0.05&42.1\p0.1&$-$2.77\p0.05&0.020\p0.001\\
0.40&$-$1.14\p0.20&42.2\p0.1&$-$2.79\p0.16&0.022\p0.002\\
\hline
\enddata
\tablenotetext{a}{\footnotesize Fixed; not fitted.}
\end{deluxetable} 


\begin {thebibliography}{dum}
\bibitem[]{}Babbedge, T.S.R. et al. 2006, \mnras, 370, 1159, 
\bibitem[]{}Brinchmann, J., Charlot, S., White, S., Tremonti, C., Kauffmann, G., Heckman, T., \& Brinkmann, J. 2004, \mnras, 351, 1151
\bibitem[]{}Cowie, L.L., Songaila, A. \& Barger, A.J. 1999, \aj, 118, 603
\bibitem[]{}Dale, D.A., Barlow, R.J., Cohen, S.A., Johnson, L.C., Kattner, S.M., Lamanna, C.A., Moore, C.A., Schuster, M.D., \& Thatcher, J.W. 2008, \aj, 135, 1412 (D08)
\bibitem[]{}Fujita, S.S, Ajiki, M., Shioya, Y. et al. 2003, \apjl, 586, L115
\bibitem[]{}Gallego, J., Zamorano, J., Arag\'on-Salamanca, A. \& Rego, M. 1995, \apjl, 455, L1 
\bibitem[]{}Gallego, J., Garc\'ia-Dab\'o, C.E., Zamorano, J., Arag\'on-Salamanca, A. \& Rego, M. 2002, \apjl, 570, L1 
\bibitem[]{}Garn, T. et al. 2010, \mnras, in press
\bibitem[]{}Geach, J.E., Smail, I., Best, P.N., Kurk, J., Casali, M., Ivison, R.J., \& Coppin, K. 2008, \mnras, 388, 1473
\bibitem[]{}Hanish, D.J. et al. 2006, \apj, 649, 150
\bibitem[]{}Hayes, M., Schaerer, D., \& \"Ostlin, G. 2010, \aap, in press
\bibitem[]{}Hippelein, H., Maier, C., Meisenheimer, K. et al. 2003, \aap, 402, 65
\bibitem[]{}Hogg, D.W., Cohen, J.G., Blandford, R., \& Pahre, M.A. 1998, \apj, 504, 622
\bibitem[]{}Hopkins, A.M. 2004, \apj, 615, 209
\bibitem[]{}Le Floc'h et al. 2005, \apj, 632, 169
\bibitem[]{}Lee, J. et al. 2010, in preparation
\bibitem[]{}Lilly, S.J., LeFevre, O., Hammer, F. \& Crampton, D. 1996, \apjl, 460, L1
\bibitem[]{}Lonsdale, C.J., et al. 2003, \pasp, 115, 897
\bibitem[]{}Ly, C. et al. 2007, \apj, 657, 738
\bibitem[]{}Ly, C. et al. 2010, in preparation
\bibitem[]{}Magnelli, B., Elbaz, D., Chary, R.R., Dickinson, M., Le Borgne, D., Frayer, D.T., \& Willmer, C.N.A. 2009, \aap, 496, 57
\bibitem[]{}Martin, D.C. et al. 2005, \apj, 521, 64
\bibitem[]{}Momcheva, I. et al. 2010, in preparation
\bibitem[]{}Moore, C.A., Dale, D.A., Barlow, R.J., Cohen, S.A., Cook, D.O., Johnson, L.C., Kattner, S.M., Lee, J.C., \& Staudaher, S.M. 2010, \aj, in press
\bibitem[]{}Morioka, T., Nakajima, A., Taniguchi, Y., Shioya, Y., Murayama, T., \& Sasaki, S.S. 2008, \pasj, 60, 1219 
\bibitem[]{}Nakamura, O., Fukugita, M., Brinkmann, J., \& Schneider, D.P. 2004, \aj, 127, 2511
\bibitem[]{}P\'erez-Gonz\'alez, P.G., Zamorano, J., Gallego, J., Arag\'on-Salamanca, A., \& Gil de Paz, A. 2003, \apj, 591, 827
\bibitem[]{}P\'erez-Gonz\'alez, P.G. et al. 2005, \apj, 630, 82
\bibitem[]{}Pierce, M.J. \& Nations, H.L. 2002, \baas, 200, 6406 
\bibitem[]{}Prescott, M., Baldy, I.K., \& James, P.A. 2009, \mnras, 397, 90
\bibitem[]{}Probst, R.G. et al. 2004, SPIE, 5492, 1716
\bibitem[]{}Rowan-Robinson, M. et al. 2008, \mnras, 386, 697
\bibitem[]{}Schechter, P. 1976, \apj, 203, 297
\bibitem[]{}Schiminovich, D. et al. 2005, \apjl, 619, L47
\bibitem[]{}Shim, H., Colbert, J., Teplitz, H., Henry, A., Malkan, M., McCarthy, P., \& Yan, L. 2009, \apj, 696, 785
\bibitem[]{}Shioya, Y. et al. 2008, \apjs, 174, 128
\bibitem[]{}Sobral, D. et al. 2009, \mnras, 398, 75
\bibitem[]{}Sullivan, M., Treyer, M.A., Ellis, R.S., Bridges, T.J., Milliard, B., \& Donas, J. 2000, \mnras, 312, 442
\bibitem[]{}Takahashi, M.I. et al. 2007, \apjs, 172, 456
\bibitem[]{}Tresse, L. \& Maddox, S.J. 1998, \apj, 495, 691
\bibitem[]{}Tresse, L., Maddox, S.J., Le Fevre, O. \& Cuby, J.G. 2002, \mnras, 337, 369
\bibitem[]{}Villar, V., Gallego, J., P\'erez-Gonz\'alez, P.G., \& Pascual, S. (2008), \apj, 677, 169
\bibitem[]{}Westra, E. \& Jones, D.H. 2008, \mnras, 383, 339
\bibitem[]{}Westra, E., Geller, M.J., Kurtz, M.J., Fabricant, D.G., \& Dell'Antonio, I. 2010, \apj, 708, 534
\bibitem[]{}Wilkins, S.M., Trentham, N., \& Hopkins, A.M. 2008, \mnras, 385, 687
\bibitem[]{}Wilson, G., Cowie, L.L., Barger, A.J., \& Burke, D.J. 2002, \aj, 124, 1258
\bibitem[]{}Yan, L., McCarthy, P.J., Freudling, W., Teplitz, H.I., Malumuth, E.M., Weymann, R.J., \& Malkan, M.A. 1999, \apjl, 519, L47
\end {thebibliography}
\end{document}